\documentclass[preprint,12pt,sort&compress]{iopjournal}

\usepackage{amssymb}
\usepackage{amsmath}
\usepackage{float} 
\usepackage{hyperref}
\usepackage{url}
\usepackage{graphicx}
\usepackage{amsmath}
\usepackage{float}

\usepackage{longtable}
\usepackage{booktabs, tabularx}
\usepackage{adjustbox}
\usepackage{multimedia}
\usepackage{footnote}
\usepackage[symbol]{footmisc}
\setfnsymbol{wiley}
\usepackage{graphicx,amsmath,amssymb,bbm,xcolor}
\usepackage{float}
\usepackage{subfig}

\def\be{\begin{equation}}
\def\lan{\left\langle}
\def\ran{\right\rangle}
\def\ee{\end{equation}}
\def\barr{\begin{array}}
\def\earr{\end{array}}

\def\dis{\displaystyle}
\def\ed{\end{document}}

\begin{document}

%\articletype{Paper}

\title{From sectorial coarse graining to extreme coarse graining of S\&P 500 correlation matrices}

\author{Manan Vyas$^{1,2,*}$,  M. Mija{\'i}l Mart{\'i}nez-Ramos$^{1, 3}$,  Parisa Majari$^{1, 4}$, Thomas H.  Seligman$^{1,2}$}

\affil{$^1$Instituto de Ciencias Físicas - Universidad Nacional Autónoma de México,  Cuernavaca, 62210,  Morelos,  México}

\affil{$^2$Centro Internacional de Ciencias AC - UNAM, Avenida Universidad 1001,  UAEM,  Cuernavaca, 62210,  Morelos,  México}

\affil{$^3$Present address: HSE University, Moscow 101000, Russia}

\affil{$^4$Present address: Tecnologico de Monterrey,  Institute of Advanced Materials for Sustainable Manufacturing, Col. Tecnológico,  Monterrey 64700,  N.L.,  México}

\affil{$^*$Author to whom any correspondence should be addressed.}

\email{manan@icf.unam.mx}

\begin{abstract}

Starting from the Pearson Correlation Matrix of stock returns and from the desire to obtain a reduced number of parameters relevant for the dynamics of a financial market,  we propose to take the idea of a sectorial matrix, which would have a large number of parameters, to the reduced picture of a real symmetric $2 \times 2$ matrix,  extreme case,  that still conserves the desirable feature that the average correlation can be one of the parameters.  This is achieved by averaging the correlation matrix over blocks created by choosing two subsets of stocks for rows and columns and averaging over each of the resulting blocks.  Averaging over these blocks, we retain the average of the correlation matrix.  We shall use a random selection for two equal block sizes as well as two specific,  hopefully relevant,  ones that do not produce equal block sizes.  The results show that one of the non-random choices has somewhat different properties,  whose meaning will have to be analyzed from an economy point of view.

\end{abstract}


\section{Introduction}

\setcounter{footnote}{0}
\renewcommand{\thefootnote}{\fnsymbol{footnote}}

A central problem of analyzing correlations of a set of time series measured or observed from the same or a related complex system is how to extract the relevant information. The term `relevant' seems to imply two aspects,  both important: for econophysics, the primary goal may be defined as understanding the dynamics of the system; for econometrics,  the primary goal is making predictions.  The two goals are not mutually exclusive,  but rather mutually supportive and have a significant overlap.  In this paper, we propose to focus on the former option and base our work on the use of states of a financial market by means of clustering of the Pearson correlation matrices of end-of-day returns taken with respect to those of the previous day. 

We choose a set of stocks, $N$ in number,  exchange-traded in the same time zone and specifically those traded in New York and as listed under the S\&P 500 index \cite{SciRep2012,  NJP2018,  Chap2023, PhyA2024, PhysScr24,  PlosOne24}.  Other techniques may be spectral analysis \cite{MSBook,  NJP2020} or principal component analysis \cite{PhyA2017, SF2025}.  All of these methods provide us with information in very high-dimensional spaces,  as is quite natural for ``big data".  Machine learning and Artificial Intelligence,  in general,  might be a more modern way to tackle the problem,  at least for econometrics,  but this may offer less access to understand dynamics, at least with current state of these tools.  

The tool of our choice is the concept of using discrete states of a financial market \cite{SciRep2012} obtained by clustering the correlation matrices into a reduced number of finite and typically fairly large sets. This yields a simple dynamics of jumps between several sets, whose number,  in our experience, varies between 4 and up to 12 in one instance \cite{SciRep2012,  NJP2018,  Chap2023, PhyA2024, PhysScr24,  PlosOne24}. The time evolution of the system in this set of states provides a transition matrix, and its properties take a central role. The time evolution can easily be represented by dots on lines corresponding to each state. 

We rely on looking for simple images that may show us some outstanding and maybe unexpected features of the data and in turn might give some insight into the underlying dynamics.  Two or three-dimensional images are certainly the preferred options.  Dimensional scaling \cite{borg2005modern} is a powerful tool and was used successfully in the context of correlation matrices \cite{ NJP2018,  Springer2019,  Chap2023} and the interpretation of these pictures shows the dominance of the largest eigenvalue or, almost equivalently, the average correlation of the entire matrix. This is not surprising as the ``state of the market" \cite{SciRep2012} is precisely defined by this parameter.  

Market states,  as introduced in \cite{SciRep2012} and expanded in \cite{NJP2018,  Chap2023, PhyA2024, PhysScr24,  PlosOne24},  deliver additional insight but the meaning of the remaining coordinates remains elusive.  On the other hand,  a recent paper \cite{PlosOne24} revealed further important properties of market states, especially the COVID state.  This state is characterized by strong reduction in correlations,  but definitively above the average correlation associated to the lowest market state.  Also, comparison with the time horizon until 2019 showed that the COVID state did not exist previously.  Indeed, the shrinking to a single market state, which essentially is independent of the label of these states, persists if ordered by the average correlation within each epoch, and the only difference we noted when moving the number of clusters from 5 up to 12 over a time horizon 2006 - 2023 is the label of this state.  The meaning of this state is still not well understood.  

In the present paper, we propose a systematic method to obtain two additional parameters. We start from the idea of forming a coarse grained (CG) matrix,  i.e.  we use the idea proposed in \cite{EPL2015} to separate the data matrix, and by consequence the correlation matrix, into a number of blocks.  An analysis of this type was performed in \cite{PhysScr24} for market sectors, which seemed to conserve much of the market state properties. Note that the CG matrix is a real symmetric matrix, i.e., it has the features of a covariance matrix. This indicates that we have $N(N+1)/2$ potentially independent matrix elements,  if we pass from the full correlation matrix to the sectorial one using the concept of CG. Then the correlation matrix is divided in sectorial blocks which are averaged over after eliminating the identities on the diagonal.  This seems to be an efficient tool of coarse graining \cite{PhysScr24} the correlation matrix.  Different kinds of coarse graining techniques have been utilized with great success in a variety of scientific problems \cite{JPC_2020,  Nat_16, NPJ_19, NPJ_25,  RA_09, PD_93,EPL_17, JASA_22,FMB_21, JCTC_22, PRX_15, PRE_07}

We plan to extend the CG method for correlation matrices to reduce the number of parameters to three due to symmetry,  using only two blocks, which is the extreme case.  One of the three parameters can be the average correlation, as this will just be the average of the CG correlation matrix. The latter is important as it essentially conserves the ``state of the market" concept.  We follow the line proposed in \cite{PhysScr24}, where the CG was performed according to the sectorial classification of the S\&P 500 into 10 blocks, determined by the number of sectors.  We will use just two blocks, which we shall choose at random but of equal size, or intuitively depending upon the intra- and inter-sectorial correlations.  

In the next section, we shall present the data and methodology we use.  In the following section, we shall try to divide each data set into two plausible subsets by selecting stronger correlations, intra- and inter-sectorial,  taken over the entire time horizon,  as well as a third one with random selections of two equally sized subsets. The last will be considered as an ensemble,  but we only sample a limited subset,  because information about outliers is not relevant for the present analysis.  We compare the three selections along with sectorial CG correlation matrix applying $k$-means clustering \cite{KMeans1, KMeans2}.  We shall see that under typical circumstances similar results are obtained at a lower computational cost,  reducing the number of parameters to three.  We finally present our conclusions and give an outlook.

{\renewcommand{\thefootnote}{\arabic{footnote}}

\section{Data and Methodology}
\label{sec1}

For the present analysis, we use as data the returns for adjusted end of day prices of  the stocks of the S\&P 500 index. The reason for this is the obvious representativity of the stocks selected as the most relevant components on the New York Stock exchange and NASDAQ.  Among these, we select all those that,  within the time horizon January 3, 2006 to August 10, 2023 ($T = 4430$ total trading days),  have no more than two consecutive trading days without a quote.  This selection  reduces the number of stocks from 500 to 322.  The corresponding stocks and their sector classification are listed in the \ref{app}. 

\begin{figure}[!h]
\centering
         \includegraphics[width=8.5cm]{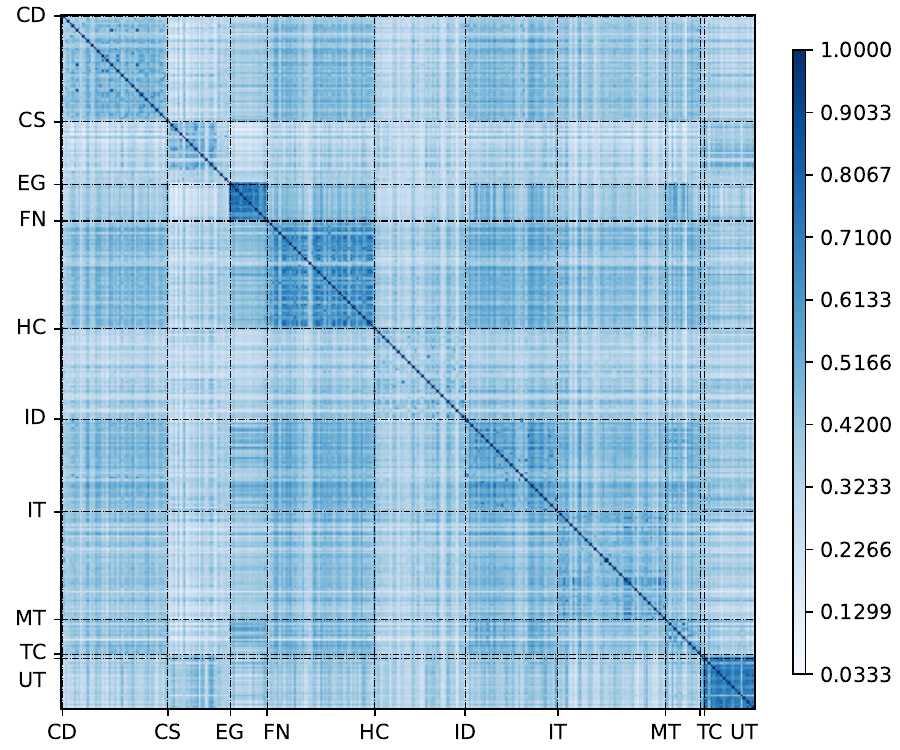}
\caption{Pearson correlation matrix $C$ defined by Eq. \eqref{eq:2} of the S\&P 500 data in a time horizon from January 3rd 2006 to August 10th 2023. Pearson correlation matrix elements are computed using logarithmic return time series of adjusted closing prices.}
\label{fig:1}
\end{figure}

\begin{figure}[!h]
\centering
         \subfloat[Choice 1]{\includegraphics[width=5cm]{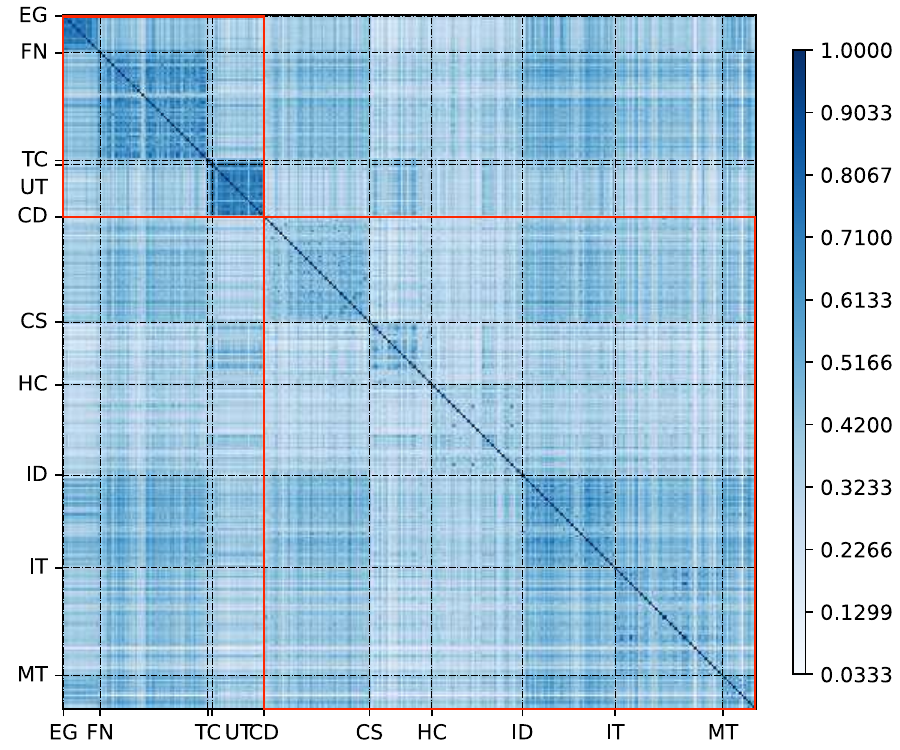}}
         \subfloat[Choice 2]{\includegraphics[width=5cm]{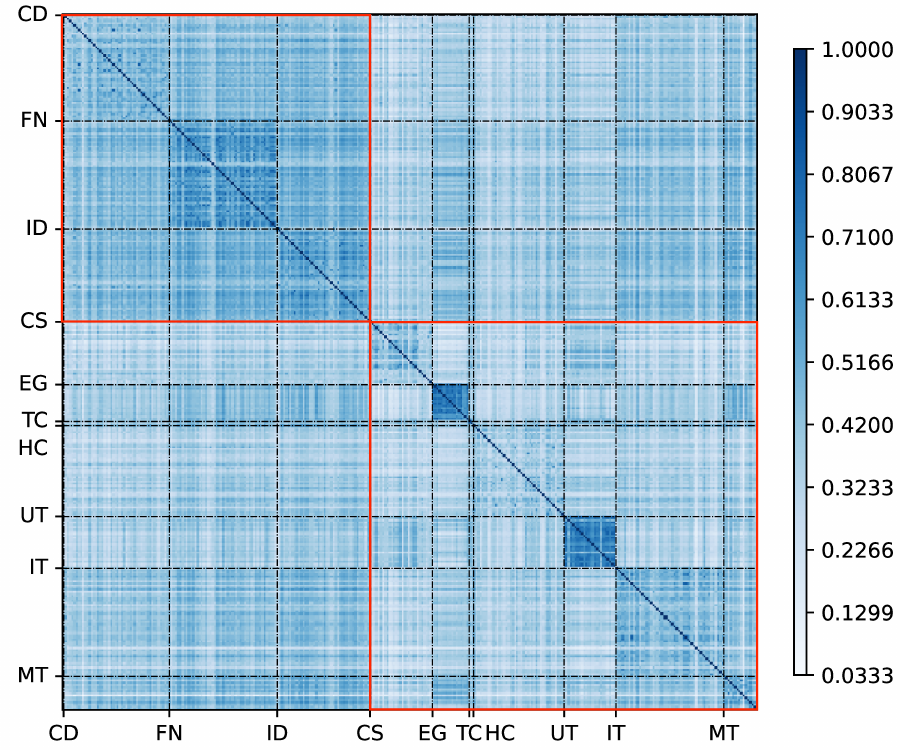}}
         \subfloat[Choice 3]{\includegraphics[width=4.85cm]{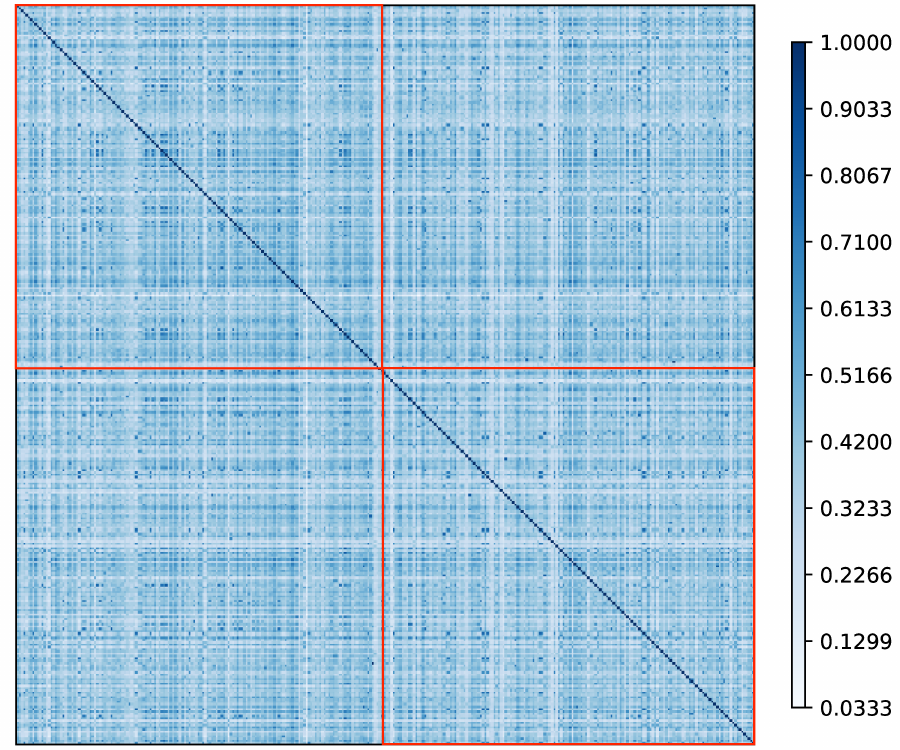}}
\caption{Choices 1, 2 and 3 for constructing the $2 \times 2$ correlation matrices.  Note that in Choice 1,  we choose sectors with strong intra-sectorial correlations (EG, FN, TC, UT) as the first block and rest of the sectors as second block; in Choice 2,  we choose sectors with strong inter-sectorial correlations (CD, FN, ID) as first block and rest of the sectors as second block; and in Choice 3, we randomly choose equal number of stocks for each block.  The choice for the blocks in each case are marked with color red.  Note that the blocks are scaled according to the number of stocks in the particular block. }
\label{fig:2}
\end{figure}

For these stocks, we find that the market states \cite{SciRep2012}, as represented by Pearson correlation matrices,  are roughly ordered according to their average correlation. This holds true,  as long as we don't choose too large a time scale for the individual epochs.  We divide the total time horizon $T$ into epochs of 20 trading days, shifted by one trading day,  and use logarithmic returns $r$ between these days as the dataset,  given the adjusted closing price $p_i(t)$ of trading day $t$ for stock $i$,
\be
r_i(t) = \log \left[ \displaystyle\frac{p_i(t)}{p_i(t-1)} \right] \;.
\label{eq:1}
\ee
For the corresponding returns, we assume zero for the days without closing quote while the return for the active trading day is computed using last active trading day.  Using these returns time series, we calculate the Pearson correlation matrix\footnote{We use the formula for the Pearson correlation matrix elements although for certain epochs and certain stocks,  the time series are strongly non-stationary.} The matrix elements of this matrix $C$ are defined as \cite{KenStu, MSBook}
\be
C_{i,j} = \dis\frac{\lan r_i \; r_j \ran - \lan r_i \ran \lan r_j \ran}{\sigma_i \, \sigma_j} \;,
\label{eq:2}
\ee
where $\sigma$ is the standard deviation of the respective return time series for the stocks. The time horizon divided into epochs of 20 trading days with one trading day shift will define our set of data matrices from which we calculate the corresponding correlation matrices.  Thus, we have $4411$ correlation matrices.

The stocks listed under the S\&P 500 market are classified into $N_s = 10$ sectors: Consumer Discretionary (CD), Consumer Staples (CS), Energy (EG), Financials (FN), Health Care (HC), Industrials (ID), Information Technology (IT), Materials (MT), Technology (TC), and Utilities (UT); see \ref{app}.  We compute and plot the Pearson correlation matrix for the total time horizon in Fig. \ref{fig:1}.  The correlations are computed using the logarithmic return time series of adjusted closing prices, as defined by Eqs. \eqref{eq:1} and \eqref{eq:2}.  One can see that there are no details about the crisis periods and all the correlations are positive due to the long time averaging performed.  In a previous work \cite{PhysScr24}, we performed averaging over the respective sectors to obtain $N_s \times N_s$ dimensional covariance matrices, which leads to significant decrease in the number of parameters to $55$ from $51681$.  In this paper, we plan to extend the analysis to $2 \times 2$ dimensional correlation matrices, which generates only three parameters.  

We can see from Fig. \ref{fig:1} that there are strong intra-sectorial correlations in sectors  EG, FN, UT and TC and strong inter-sectorial correlations in sectors CD, FN, and IT.  Thus, we group EG, FN, UT and TC into one block and the rest of the sectors into the second block as choice 1; we group CD, FN, and IT into one block and the rest of the sectors into the second block as choice 2; and we choose equal number of stocks randomly for the two blocks as choice 3.  Figure \ref{fig:2} explains these choices.  Note that the blocks are scaled according to the number of stocks in the particular block.  Then, we sum all the correlation matrix elements inside each block (excluding the self-correlations) and divide by the total number of correlation matrix elements in the block resulting in a $2 \times 2$ dimensional extreme coarse grained (ECG) correlation matrix for each epoch.  Note that coarse graining beyond the $2 \times 2$ block matrices, we obtain the average correlation and thus, the terminology {\it extreme coarse graining}.  We also make a comparison with the $N_s \times N_s$ dimensional sectorial coarse grained (CG) correlation matrix.  Considering each intra- and inter-sectorial blocks, we sum all the correlation matrix elements inside each block (excluding the self-correlations) and divide by the total number of correlation matrix elements in the block to obtain the $10 \times 10$ sectorial CG correlation matrix \cite{PhysScr24}.

\begin{figure}[!h]
\centering
        \subfloat[Sectorial CG]{\includegraphics[width=14.5cm]{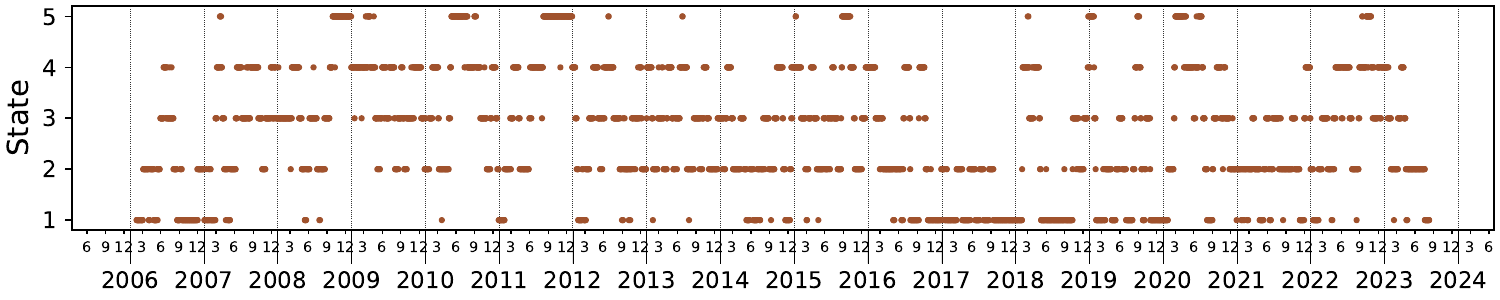}} \\
         \subfloat[Choice 1  ECG]{\includegraphics[width=14.5cm]{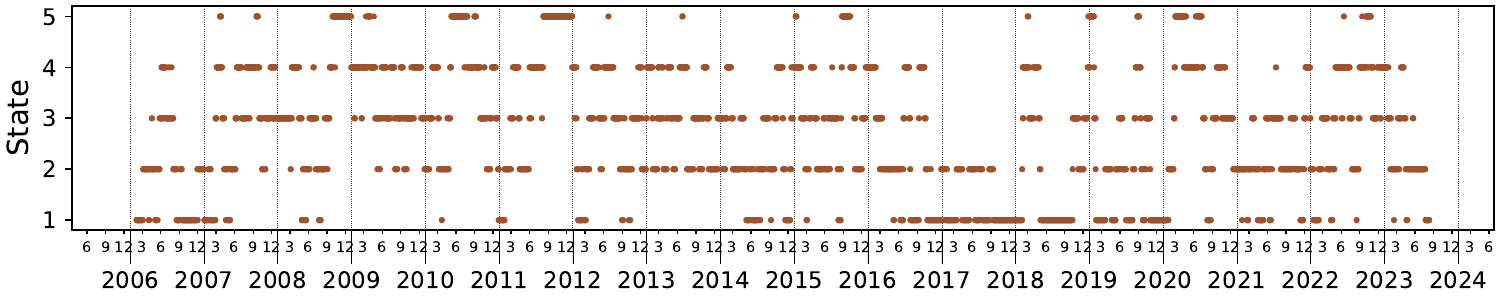}} \\
         \subfloat[Choice 2  ECG]{\includegraphics[width=14.5cm]{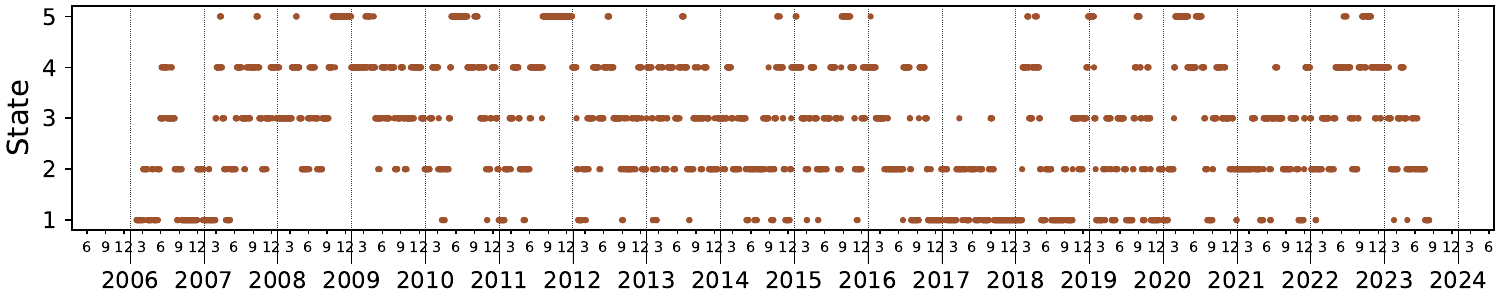}} \\
         \subfloat[Choice 3  ECG]{\includegraphics[width=14.5cm]{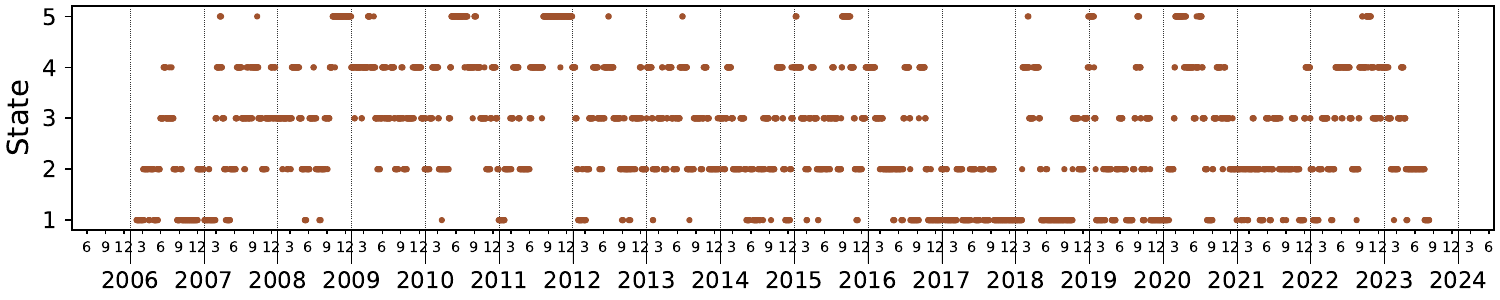}}
\caption{Time evolution of market states of the S\&P 500 data using (a) CG and (b)-(d)  ECG Pearson correlation matrix $C$ defined by Eq. \eqref{eq:2} in a time horizon from January 3rd 2006 to August 10th 2023 with an epoch of 20 trading days. Pearson correlation matrix elements are computed using logarithmic return time series of adjusted closing prices.  The market states are arranged in order of increasing average correlations.  The average correlations for the states are (a) 0.1586, 0.2654, 0.3734, 0.4842, 0.6462; (b) 0.1496, 0.2613, 0.3706, 0.4838, 0.6465; (c) 0.165, 0.269, 0.372, 0.484, 0.643; and (d) 0.1493,  0.2594, 0.3704, 0.4809,  0.6429; respectively.  The Pearson correlation coefficients among all combinations between Figs. (a)-(d) are above $0.92$.}
\label{fig:3}
\end{figure}

\section{Results and discussion}
\label{sec2}

Using the correlation matrices for each epoch shifted by one day,  we first calculate the CG and ECG matrices for the three choices explained in Sec. \ref{sec1}. Using $k$-means clustering \cite{KMeans1, KMeans2}, we group these in five states and the time evolution of S\&P 500 market is as shown in Fig. \ref{fig:3}.  The first thing we note is that the COVID state \cite{PlosOne24} does not appear for both CG and ECG matrices.  We have verified that the COVID state is very sensitive to multi-dimensional scaling \cite{borg2005modern},  CG \cite{PhysScr24},  and Power-Map technique \cite{Guhr_2003, THS_2013}  and it does not appear as correlations are very weak and the state is fragile.  We also notice a vertical gap in year 2017 that seems to be a signature of a very calm period and hence,  CG and ECG can identify calm periods.  Importantly, this gap was present in the analysis using $322 \times 322$ dimensional Pearson correlation matrices \cite{PlosOne24}.  Earlier work \cite{NJP2018} also shows a similar gap around the year 2000 and it is significant that this gap survives while using CG and  ECG.  Importantly, CG and three choices with  ECG results are very similar and hence, provides an effective way of parameter reduction without losing very important information. 

\begin{figure}[!h]
 \centering
        \includegraphics[width=14.5cm]{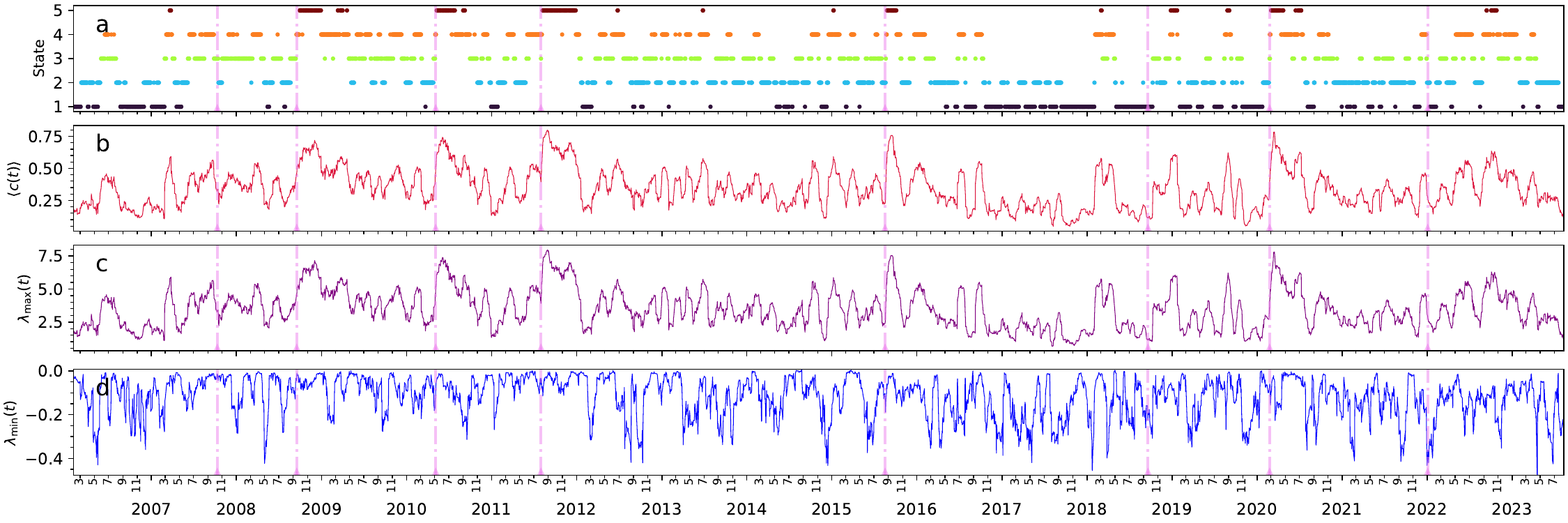}
\caption{ {\bf (a)} Time evolution of market states of the S\&P 500 data using CG Pearson correlation matrix $C$ defined by Eq. \eqref{eq:2} in a time horizon from January 3rd 2006 to August 10th 2023 with an epoch of 20 trading days for CG Pearson correlation matrices.  Each state is represented by a different color and dashed horizontal lines indicate the dates of stock market crashes; see Table \ref{tab1} for details.  Time evolution of {\bf (b)} average correlations, {\bf (c)} largest eigenvalue, and {\bf (d)} smallest eigenvalue.  The Pearson correlation coefficients between average correlation and $\lambda_{max}$ is 0.998,  average correlation and $\lambda_{min}$ is 0.364,  $\lambda_{max}$ and $\lambda_{min}$ is 0.350.}
\label{fig:4}
\end{figure}

\begin{figure}[!h]
 \centering
        \includegraphics[width=14.5cm]{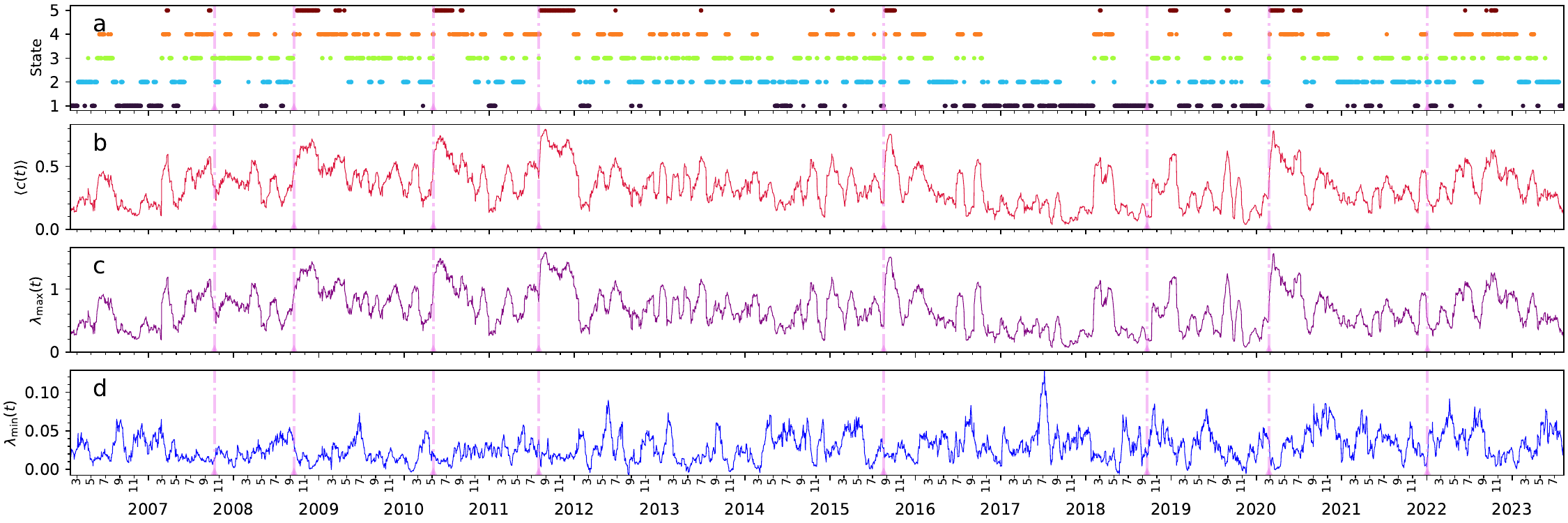}
\caption{{\bf (a)} Time evolution of market states of the S\&P 500 data using  ECG Pearson correlation matrix $C$ defined by Eq. \eqref{eq:2} in a time horizon from January 3rd 2006 to August 10th 2023 with an epoch of 20 trading days for Choice 1,  same as in Fig. \ref{fig:2}.  Each state is represented by a different color and dashed horizontal lines indicate the dates of stock market crashes; see Table \ref{tab1} for details.  Time evolution of {\bf (b)} average correlations, {\bf (c)} largest eigenvalue, and {\bf (d)} smallest eigenvalue.  The Pearson correlation coefficients between average correlations and $\lambda_{max}$ is 0.999,  average correlations and $\lambda_{min}$ is -0.311,  $\lambda_{max}$ and $\lambda_{min}$ is -0.328.}
\label{fig:5}
\end{figure}

\begin{table}[h]
\centering
\begin{tabular}{|c|c|}
    \toprule
    \textbf{Date (DD/MM/YYYY)} & \textbf{Name} \\
    \midrule
11/10/2007 & United States bear market \\ \hline
16/09/2008 & Financial crisis of 2007–2008 \\ \hline
06/05/2010  & 2010 flash crash\\ \hline
01/08/2011 & August 2011 stock markets fall\\ \hline
18/08/2015 & 2015–2016 stock market selloff \\ \hline
20/09/2018 &  Cryptocurrency crash \\ \hline
24/02/2020  &  COVID19 crash \\ \hline
03/01/2022  &  2022 stock market decline \\ \hline
\midrule
\end{tabular}  
\caption{List of crashes considered}
\label{tab1}
\end{table}

We also analyze the behavior of average correlations and the two eigenvalues $(\lambda_{min}, \lambda_{max})$ focusing on the stock market crashes listed in Table \ref{tab1}. Around these crisis periods, the market is in the state with largest average correlations. Also, the average correlations are strongly positively correlated with the largest eigenvalue.  Notably, the largest eigenvalue represents the `State of the Market' \cite{SciRep2012} and the CG technique \cite{PhysScr24} preserves this relevant feature as can be seen from Fig. \ref{fig:4}. The Pearson correlation coefficients between average correlation and $\lambda_{max}$ is 0.998,  average correlation and $\lambda_{min}$ is 0.364,  $\lambda_{max}$ and $\lambda_{min}$ is 0.350.  Using the three different choices for  ECG, explained in Fig. \ref{fig:2}, we perform similar analysis and the respective results are presented in Figs.  \ref{fig:5}-\ref{fig:7}.  The results are similar to those obtained with CG, however, the Pearson correlation coefficient between average correlations (or, equivalently $\lambda_{max}$) is negative for the  ECG choices. 

\begin{figure}[!h]
 \centering        
\includegraphics[width=14.5cm]{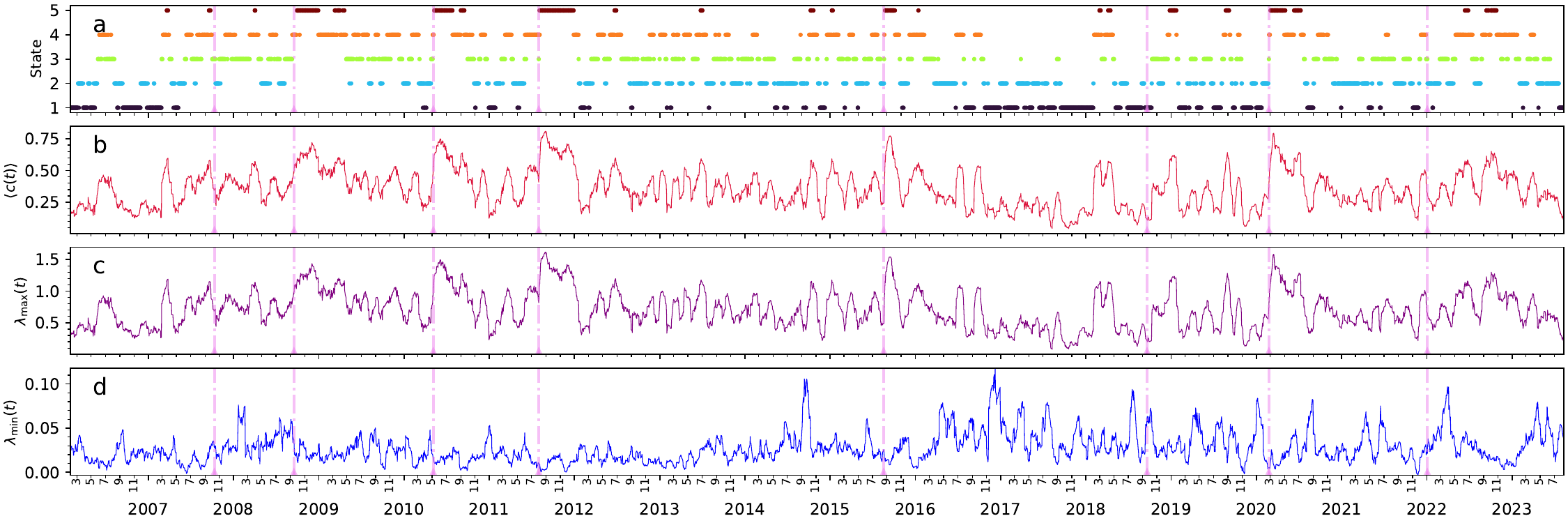}
\caption{{\bf (a)} Time evolution of market states of the S\&P 500 data using  ECG Pearson correlation matrix $C$ defined by Eq. \eqref{eq:2} in a time horizon from January 3rd 2006 to August 10th 2023 with an epoch of 20 trading days for Choice 2,  same as in Fig. \ref{fig:2}.  Each state is represented by a different color and dashed horizontal lines indicate the dates of stock market crashes; see Table \ref{tab1} for details.  Time evolution of {\bf (b)} average correlations, {\bf (c)} largest eigenvalue, and {\bf (d)} smallest eigenvalue.  The Pearson correlation coefficients between average correlations and $\lambda_{max}$ is 0.999,  average correlations and $\lambda_{min}$ is -0.377,  $\lambda_{max}$ and $\lambda_{min}$ is -0.389.}
\label{fig:6}

\end{figure}
\begin{figure}[!h]
\centering
         \includegraphics[width=14.5cm]{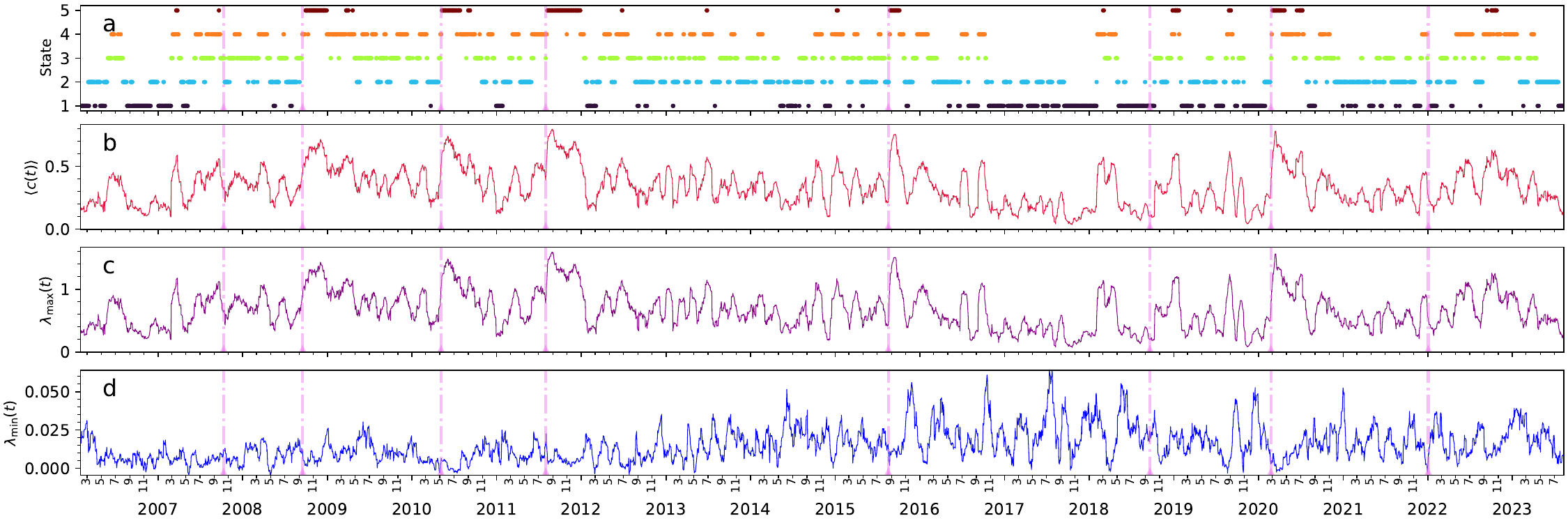}
\caption{{\bf (a)} Time evolution of market states of the S\&P 500 data using  ECG Pearson correlation matrix $C$ defined by Eq. \eqref{eq:2} in a time horizon from January 3rd 2006 to August 10th 2023 with an epoch of 20 trading days for Choice 3,  same as in Fig. \ref{fig:2}.  Each state is represented by a different color and dashed horizontal lines indicate the dates of stock market crashes; see Table \ref{tab1} for details.  Time evolution of {\bf (b)} average correlations, {\bf (c)} largest eigenvalue, and {\bf (d)} smallest eigenvalue.  The Pearson correlation coefficients between average correlations and $\lambda_{max}$ is 0.999,  average correlations and $\lambda_{min}$ is -0.348,  $\lambda_{max}$ and $\lambda_{min}$ is -0.358.}
\label{fig:7}
\end{figure}

\begin{figure}[!h]
\centering
         \subfloat[Sectorial CG]{\includegraphics[width=14.5cm]{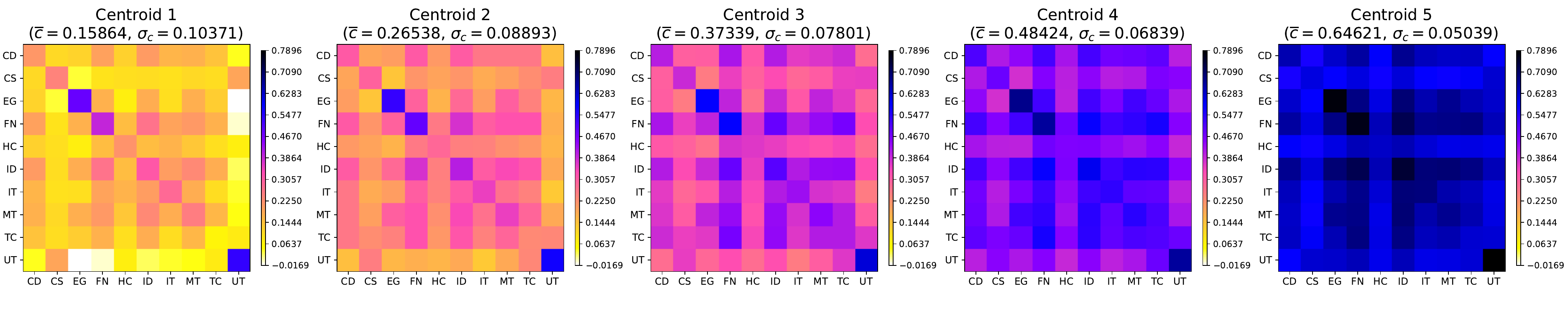}} \\
         \subfloat[Choice 1  ECG]{\includegraphics[width=14.5cm]{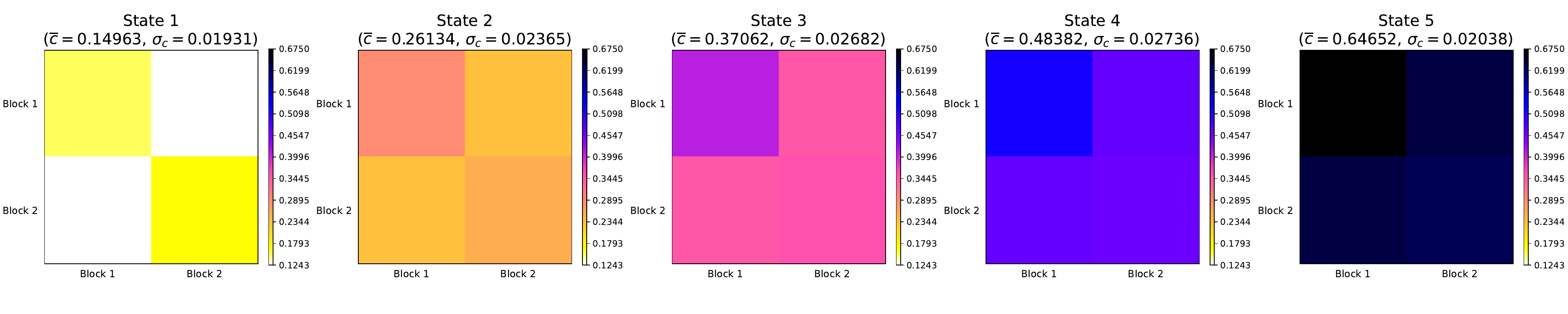}} \\
         \subfloat[Choice 2  ECG]{\includegraphics[width=14.5cm]{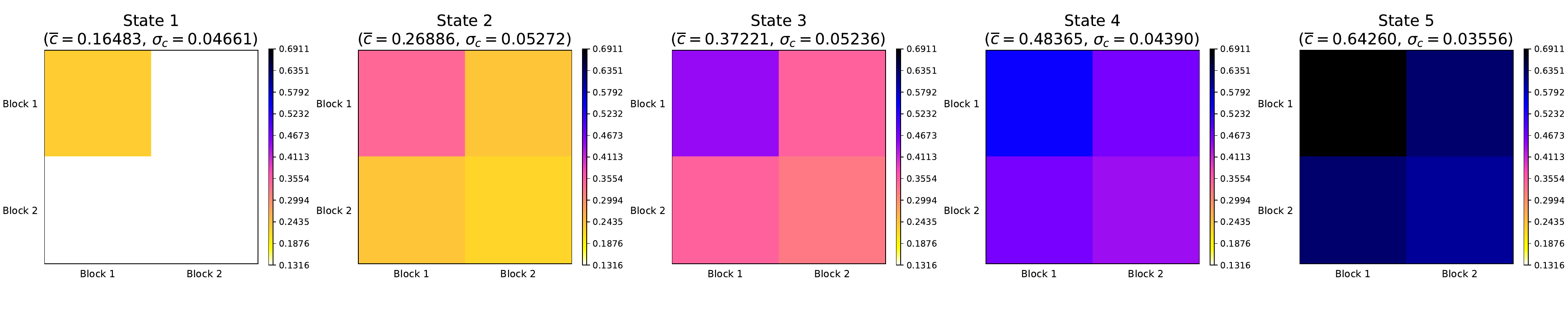}}\\
         \subfloat[Choice 3  ECG]{\includegraphics[width=14.5cm]{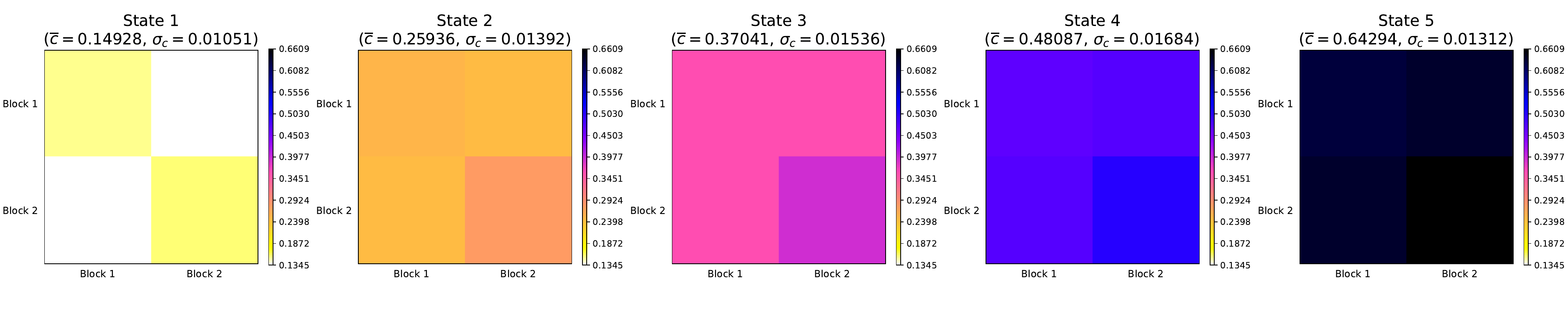}}
\caption{Average correlation matrices for the five market states, shown in Fig. \ref{fig:3},  corresponding to (a) Sectorial CG, (b)  ECG Choice 1, (c)  ECG Choice 2, and (d)  ECG Choice 3, Pearson correlation matrices.  Note that the market states are arranged according to increasing value of average correlations $\overline{c}$.}
\label{fig:12}
\end{figure}

\begin{figure}[!h]
\centering
        \subfloat[Sectorial CG]{\includegraphics[width=3.5cm]{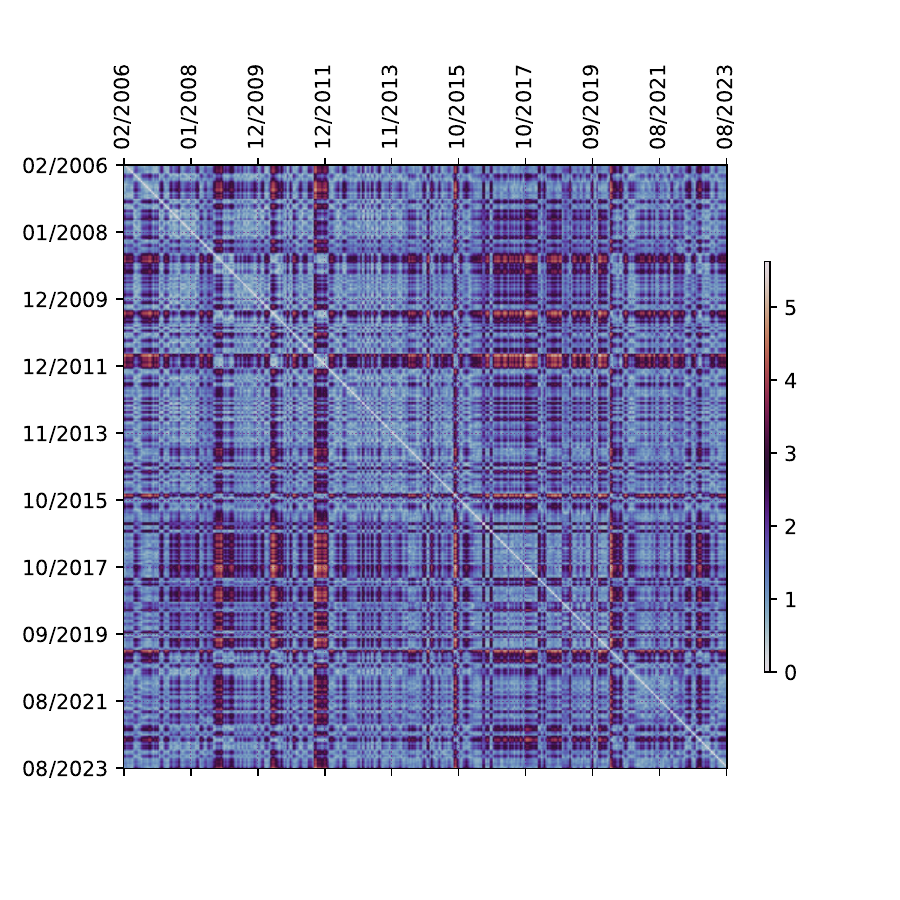}} 
         \subfloat[Choice 1  ECG]{\includegraphics[width=3.5cm]{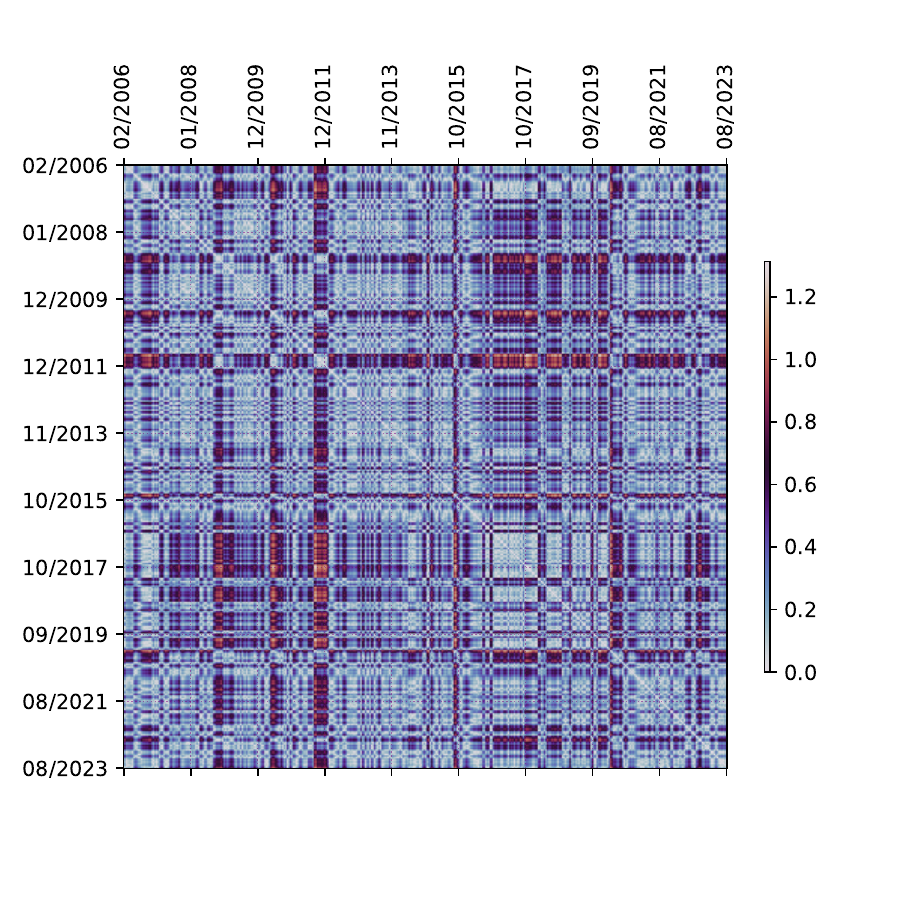}}
         \subfloat[Choice 2  ECG]{\includegraphics[width=3.5cm]{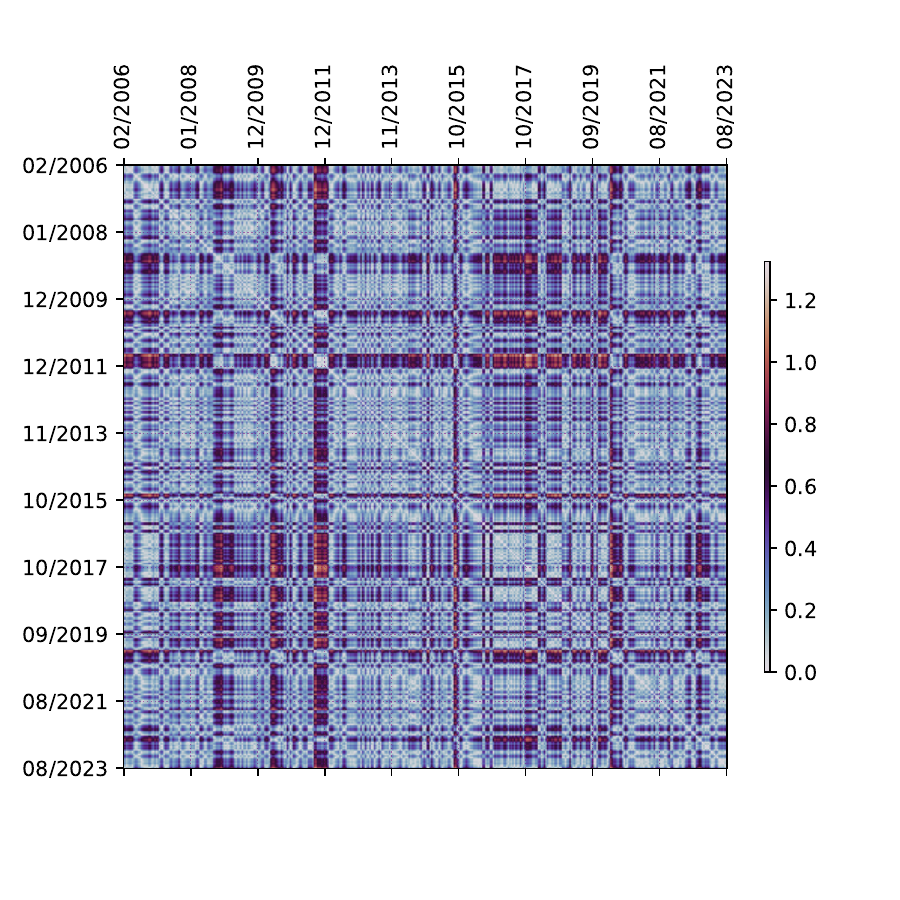}}
         \subfloat[Choice 3  ECG]{\includegraphics[width=3.5cm]{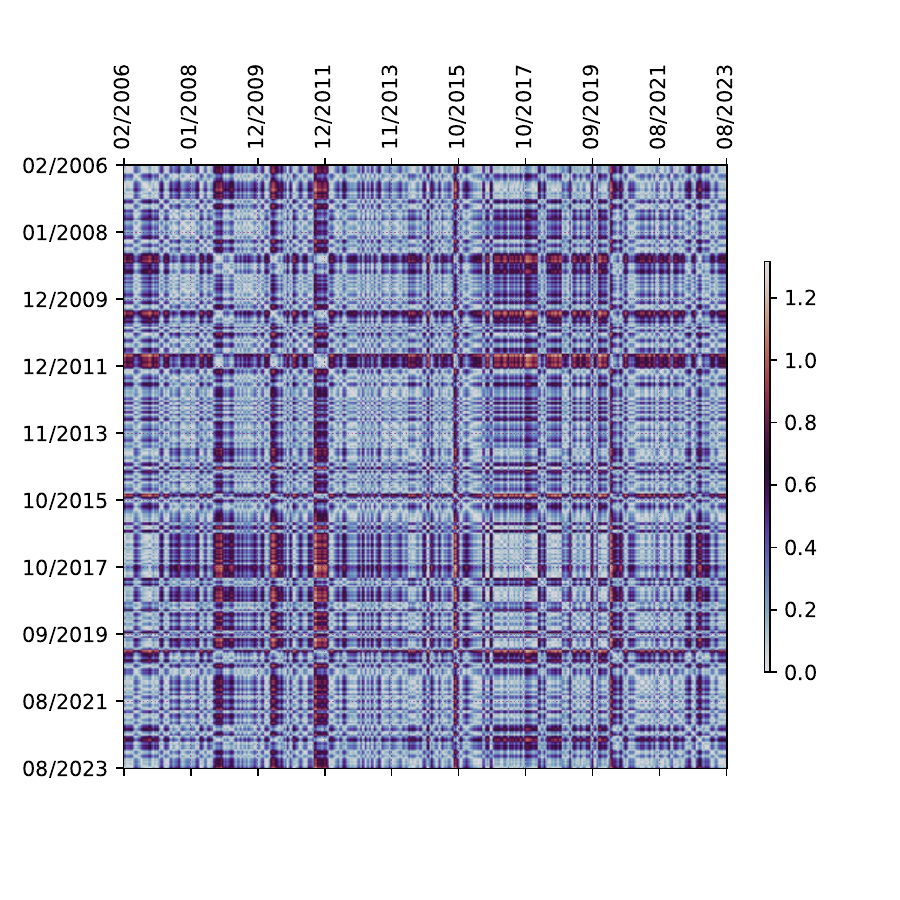}}
\caption{Similarity matrices corresponding to (a) Sectorial CG, (b)  ECG Choice 1, (c)  ECG Choice 2, and (d)  ECG Choice 3 for S\&P 500 in Fig.  \ref{fig:3}. }
\label{fig:13}
\end{figure}

\begin{figure}[!h]
\centering
        \subfloat[Sectorial CG]{\includegraphics[width=3.5cm]{
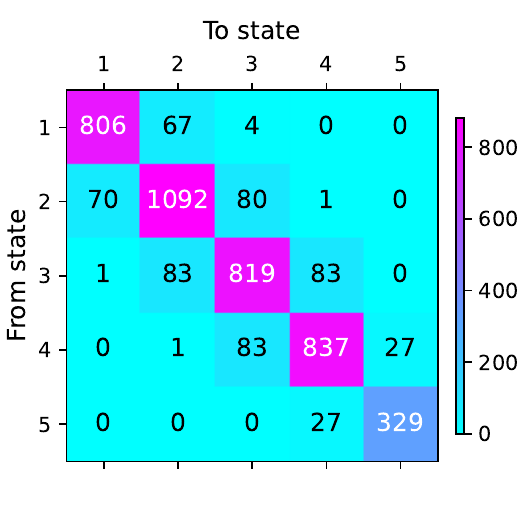}}
         \subfloat[Choice 1 ECG]{\includegraphics[width=3.5cm]{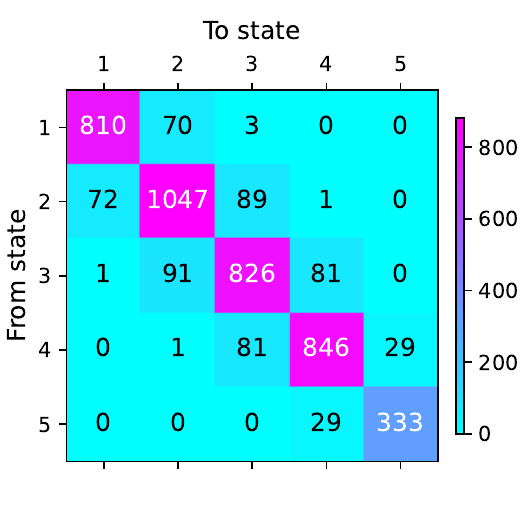}}
         \subfloat[Choice 2 ECG]{\includegraphics[width=3.5cm]{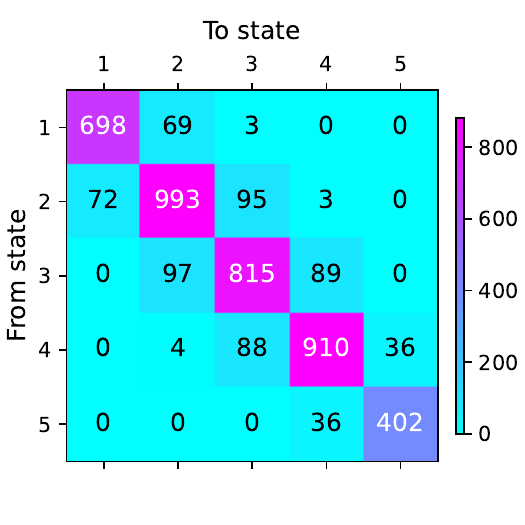}}
         \subfloat[Choice 3 ECG]{\includegraphics[width=3.5cm]{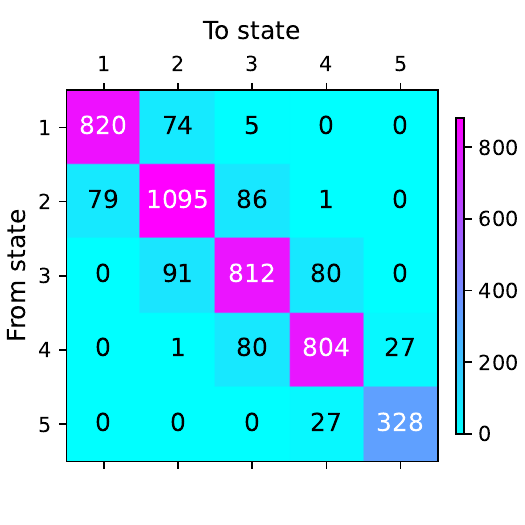}}
\caption{Transition matrices showing transitions between different market states for S\&P 500 in Fig.  \ref{fig:3} obtained with (a) Sectorial CG, (b)  ECG Choice 1, (c)  ECG Choice 2, and (d)  ECG Choice 3, Pearson correlation matrices.  The transition matrices are near-tridiagonal.  Also, the necessary criterion for Markovianity given in Eq. (2) of \cite{NJP2018} is fulfilled for both.  The equilibrium distributions corresponding to (a)-(d) respectively are (0.1989,  0.2819,  0.2236,  0.2150 ,  0.0807), (0.2002, 0.2741, 0.2265, 0.217, 0.0821), (0.1746, 0.2637, 0.2270, 0.2354, 0.0993) and (0.2039, 0.2859, 0.2229, 0.2068, 0.0805) respectively.}
\label{fig:14}
\end{figure}

\begin{figure}[!h]
\centering
         \subfloat[Choice 1 ECG]{\includegraphics[width=14.5cm]{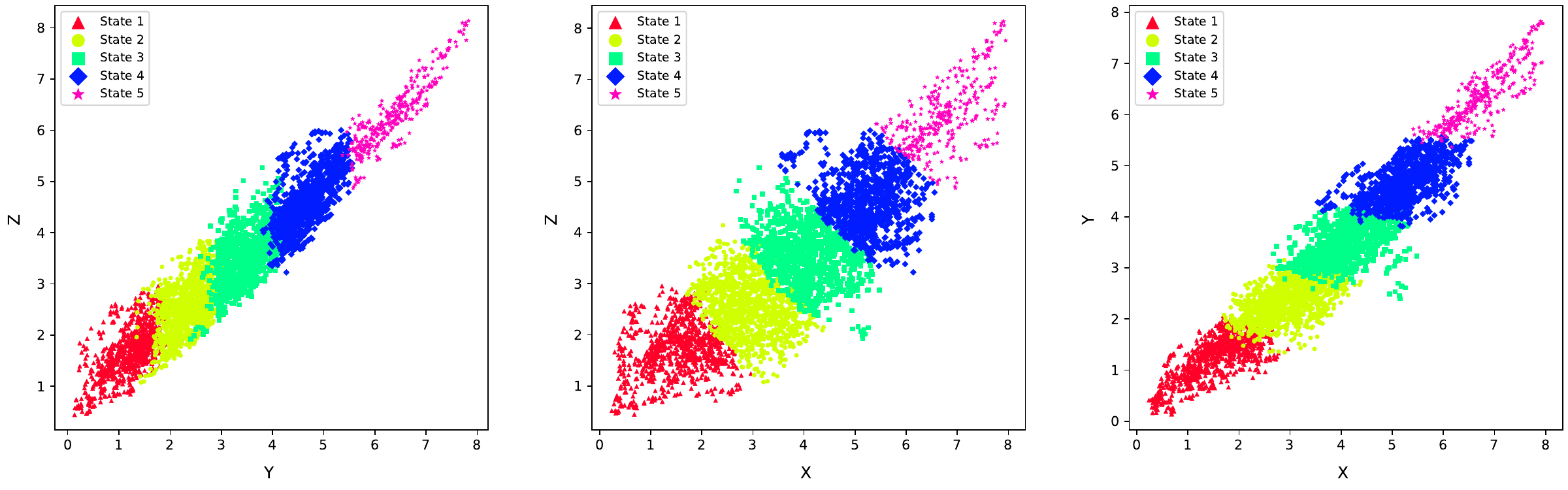}} \\
         \subfloat[Choice 2 ECG]{\includegraphics[width=14.5cm]{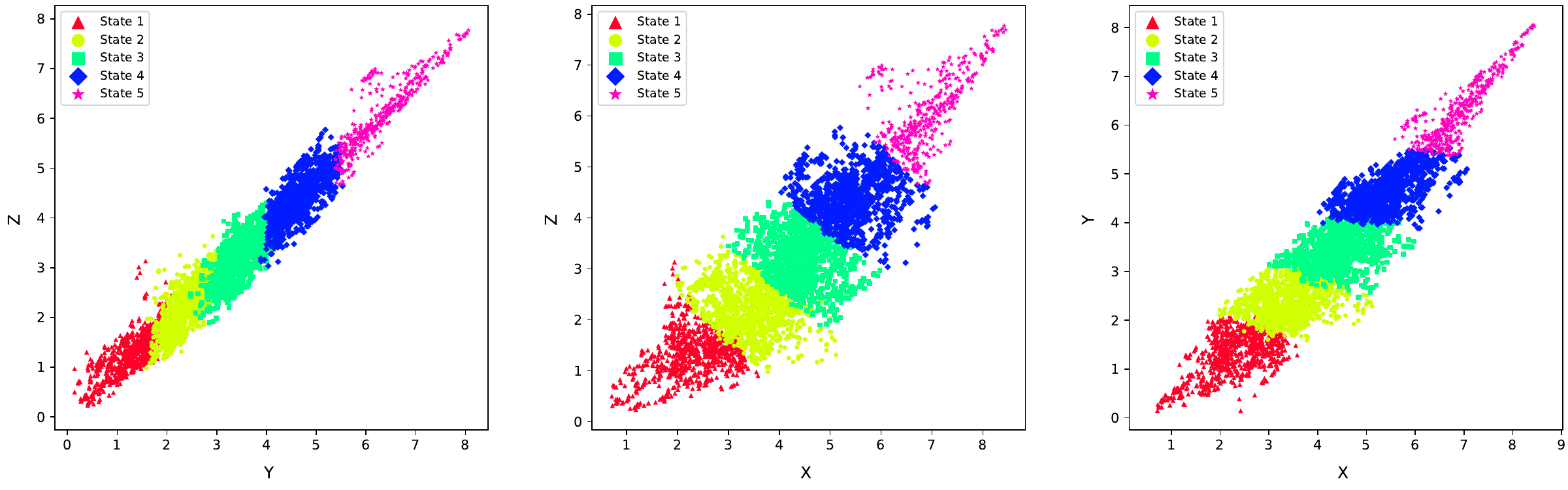}} \\
         \subfloat[Choice 3 ECG]{\includegraphics[width=14.5cm]{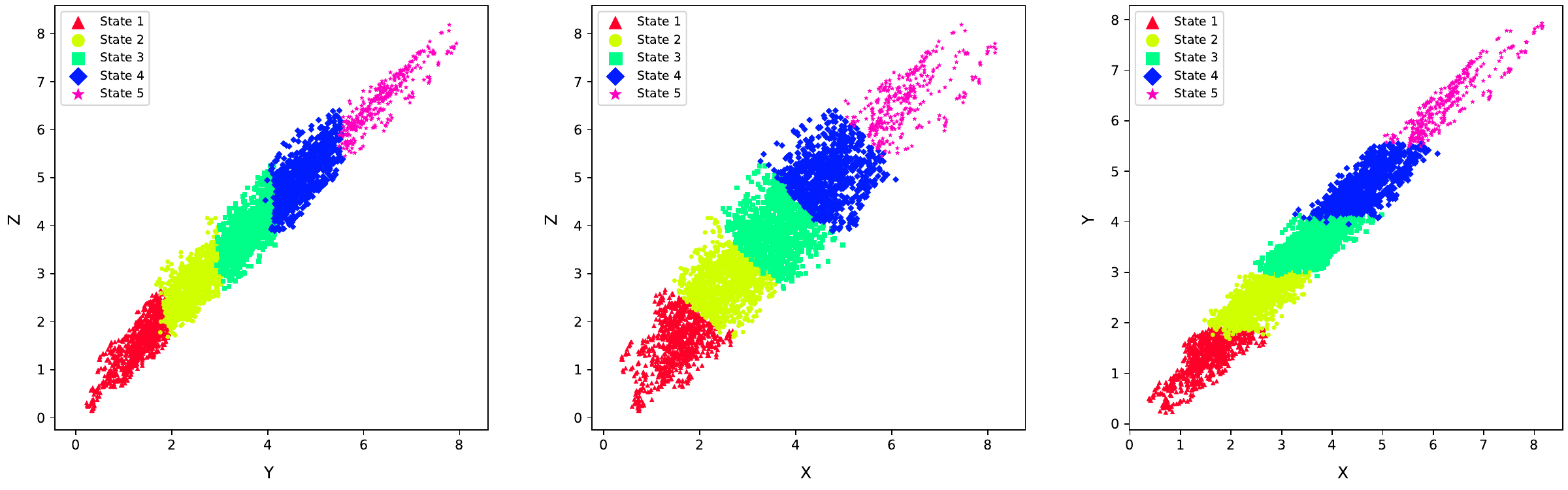}}
\caption{Correlation matrix elements $(x,y,z)$ for different time epochs, color coded according to the market states obtained using $k$-means clustering in Fig. \ref{fig:3},  for (a) ECG Choice 1, (b)  ECG Choice 2, and (c) ECG Choice 3.  Note that $y$ is the off-diagonal correlation matrix element and no dimensional reduction technique has been used.}
\label{fig:15}
\end{figure}

The variances for the CG matrices decrease, as expected, around the crisis periods. The skewness $\gamma_1$ is always $\ge 0$, except for the COVID period where $\gamma_1 < 0$. The kurtosis is always $\gamma_2 \ge 0$.  With the three  ECG choices, the variance shows a similar trend as with the CG matrices. However,  it is difficult to infer anything conclusive about $\gamma_1$ and $\gamma_2$ for the three  ECG choices due to large fluctuations. The fluctuations in $\gamma_1$ for Choice 2 are relatively smaller in comparison to those obtained with Choices 1 and 3. 

Figure \ref{fig:12}(a)-(d) respectively shows the average correlation matrices corresponding to each state for sectorial CG,  ECG Choice 1,  ECG Choice 2 and  ECG Choice 3 Pearson correlation matrices.  The values of respective mean $\overline{c}$ and standard deviation $\sigma_c$ are as indicated in the figure.  The figures for the three  ECG choices are similar, re-enforcing the idea that this technique preserves the essential features.  We also compare the corresponding similarity matrices in Fig. \ref{fig:13} and observe that the information about the crisis periods is preserved using both the CG and  ECG techniques.

Transition matrices for sectorial CG and the three  ECG choices are near-tridiagonal as shown in Fig. \ref{fig:14}.  Notably, we also preserve the feature that there are only transitions from state 4 (near-critical state) to state 5 (critical state).  There are no transitions from states 1-3 to state 5.  In the block between states 4 and 5,  ECG Choice 3 has more states while  ECG Choices 1 and 3 are more similar.

For the three ECG choices, we have $2 \times 2$ block correlation matrices and thus, three matrix elements $(x, y, z)$.  Without any dimensional reduction \cite{borg2005modern}, we plot these correlation matrix elements $(y,z)$, $(x,z)$ and $(x,y)$ in Fig. \ref{fig:15} for the three choices explained in Fig. \ref{fig:2}.  Note that $y$ is the off-diagonal correlation matrix element.  These are very similar to the plots obtained using full correlation matrices of size $322 \times 322$ after performing multi-dimensional scaling to three dimensions \cite{PlosOne24}.  Spread in the $(x,z)$ and $(x,y)$ plots is the smallest for the ECG Choice 3 compared to the other two choices.  One can possibly analyze various linear combinations between these variables as well. 

\section{Conclusions and future outlook}
\label{sec3} 

In the pursuit to find a reduced set of parameters to identify the dynamics seen for discrete states for financial markets \cite{PlosOne24},  it has been shown \cite{PhysScr24} that the reduction using coarse-grained correlation according to the sectorial classification of the S\&P 500 yields a great reduction of the number of parameters to 55, which still was unwieldy.  In the attempt to reduce the coarse-grained matrix to $2 \times 2$ correlation matrices, extreme case,  with 3 parameters respectively,  which would be manageable and would allow fairly simple illustrations.  We have divided each data set into two plausible subsets by selecting stronger correlations, intra- and inter-sectorial,  considering the entire time horizon,  as well as a third one with random selections of two equally sized subsets. The last will be considered as an ensemble,  but we only sample a limited subset.  We have not addressed this statistical problem numerically,  as with hundreds of stocks,  it is exponentially big.  But only ran 1000 out of a sample of $322!$ possible random arrangements taken over the entire time horizon at will.  Specific situations thus can easily deviate to different degrees.  Yet one important fact must not be forgotten: The average correlation is conserved and it is highly correlated to the largest eigenvalue.  If the latter is large,  the ``space of the correlation matrices" shrinks,  and thus an important feature,  namely high correlations usually associated with crisis, are reflected by a parameter,  which is independent of any permutation.  This also gives a hint why low correlations associated with very quiet stock markets are not too well reflected,  and the non-appearance of the ``COVID anomaly" \cite{PlosOne24} is no great surprise.  Indeed,  this already happens with the sectorial coarse-graining.  Although the choice of the time horizon is essential for our conclusions but maybe not very restrictive.  All the three choices yield similar results which are obtained at a lower computational cost,  reducing the number of parameters to three.  Thus, we conclude that a three-dimensional parameter space explains the evolution of market states in a three-dimensional picture to a large extent,  though the COVID anomaly may require a more refined treatment.  The ECG technique introduced in this paper allows to focus on a minimal subset of parameters which is useful to identify underlying structures in complex systems in diverse areas like physics, finance, molecular dynamics, computational biology, information theory, and so on \cite{Guhr_2022, Nat_16, NPJ_19, NPJ_25,  RA_09, PD_93,EPL_17, JASA_22,FMB_21, JCTC_22, PRX_15, PRE_07}.  These CG and ECG results are largely similar with hierarchical clustering as well \cite{Ma-23}. 

As far as future work is concerned,  these ideas must be extended to other markets and other options for parameter choices must be considered.  One must also be aware that noise-suppression techniques such as the Power-Map \cite{Guhr_2003, THS_2013} as well as soft windowing choices including wavelet techniques, can be useful.  Yet we have found that the powermap, as well as dimensional scaling, both affect the discretization implied in the concept of market states, destroy some features like COVID anomaly.  

\data

{Data is available in figshare repository  \url{https://doi.org/10.6084/m9.figshare.25219880.v1}}; downloaded from \url{https://finance.yahoo.com/}.

\ack

The authors are grateful to Thomas Guhr,  Francois Leyvraz,  Anirban Chakraborti,  and Hirdesh K. Pharasi for useful discussions.  

\funding

We acknowledge financial support from Universidad Nacional Autónoma de México (UNAM) Dirección General de Asuntos del Personal Académico (DGAPA) Programa de Apoyo a Proyectos de Investigación e Innovación Tecnológica (PAPIIT) Project AG101122.  The funders had no role in study design, data collection and analysis, decision to publish, or preparation of the manuscript.

\newpage

\appendix
\renewcommand{\thesection}{Appendix}
\section{List of the 322 stocks analyzed with their sectorial classification}
\label{app} 

{\tiny
%\begin{adjustbox}{width=1\textwidth}
\begin{longtable}[!p]{lll}
%\begin{tabular}{lllll}
\toprule
\textbf{Sector} & \textbf{Ticker} & \textbf{Name}  \\
\midrule
Basic Materials & APD & Air Products and Chemicals, Inc. \\
Basic Materials & CF & CF Industries Holdings, Inc. \\
Basic Materials & ECL & Ecolab Inc. \\
Basic Materials & FCX & Freeport-McMoRan Inc. \\
Basic Materials & FMC & FMC Corporation \\
Basic Materials & IFF & International Flavors \& Fragrances Inc. \\
Basic Materials & MOS & The Mosaic Company \\
Basic Materials & NEM & Newmont Corporation \\
Basic Materials & NUE & Nucor Corporation \\
Basic Materials & PPG & PPG Industries, Inc. \\
Basic Materials & SHW & The Sherwin-Williams Company \\
Basic Materials & VMC & Vulcan Materials Company \\
Communication Services & ATVI & Activision Blizzard, Inc. \\
Communication Services & CMCSA & Comcast Corporation \\
Communication Services & DISH & DISH Network Corporation \\
Communication Services & EA & Electronic Arts Inc. \\
Communication Services & GOOG & Alphabet Inc. \\
Communication Services & GOOGL & Alphabet Inc. \\
Communication Services & IPG & The Interpublic Group of Companies, Inc. \\
Communication Services & NFLX & Netflix, Inc. \\
Communication Services & OMC & Omnicom Group Inc. \\
Communication Services & T & AT\&T Inc. \\
Communication Services & TTWO & Take-Two Interactive Software, Inc. \\
Communication Services & VZ & Verizon Communications Inc. \\
Consumer Cyclical & AAP & Advance Auto Parts, Inc. \\
Consumer Cyclical & AMZN & Amazon.com, Inc. \\
Consumer Cyclical & AVY & Avery Dennison Corporation \\
Consumer Cyclical & AZO & AutoZone, Inc. \\
Consumer Cyclical & BBY & Best Buy Co., Inc. \\
Consumer Cyclical & BKNG & Booking Holdings Inc. \\
Consumer Cyclical & CCL & Carnival Corporation \& plc \\
Consumer Cyclical & DHI & D.R. Horton, Inc. \\
Consumer Cyclical & EBAY & eBay Inc. \\
Consumer Cyclical & EXPE & Expedia Group, Inc. \\
Consumer Cyclical & F & Ford Motor Company \\
Consumer Cyclical & GPC & Genuine Parts Company \\
Consumer Cyclical & GPS & The Gap, Inc. \\
Consumer Cyclical & HAS & Hasbro, Inc. \\
Consumer Cyclical & HD & The Home Depot, Inc. \\
Consumer Cyclical & HOG & Harley-Davidson, Inc. \\
Consumer Cyclical & HRB & H\&R Block, Inc. \\
Consumer Cyclical & IP & International Paper Company \\
Consumer Cyclical & JWN & Nordstrom, Inc. \\
Consumer Cyclical & KMX & CarMax, Inc. \\
Consumer Cyclical & KSS & Kohl's Corporation \\
Consumer Cyclical & LEG & Leggett \& Platt, Incorporated \\
Consumer Cyclical & LEN & Lennar Corporation \\
Consumer Cyclical & LKQ & LKQ Corporation \\
Consumer Cyclical & LOW & Lowe's Companies, Inc. \\
Consumer Cyclical & M & Macy's, Inc. \\
Consumer Cyclical & MAR & Marriott International, Inc. \\
Consumer Cyclical & MCD & McDonald's Corporation \\
Consumer Cyclical & MGM & MGM Resorts International \\
Consumer Cyclical & MHK & Mohawk Industries, Inc. \\
Consumer Cyclical & NKE & NIKE, Inc. \\
Consumer Cyclical & ORLY & O'Reilly Automotive, Inc. \\
Consumer Cyclical & PHM & PulteGroup, Inc. \\
Consumer Cyclical & PKG & Packaging Corporation of America \\
Consumer Cyclical & PVH & PVH Corp. \\
Consumer Cyclical & RL & Ralph Lauren Corporation \\
Consumer Cyclical & ROST & Ross Stores, Inc. \\
Consumer Cyclical & SBUX & Starbucks Corporation \\
Consumer Cyclical & SEE & Sealed Air Corporation \\
Consumer Cyclical & TJX & The TJX Companies, Inc. \\
Consumer Cyclical & TPR & Tapestry, Inc. \\
Consumer Cyclical & UAA & Under Armour, Inc. \\
Consumer Cyclical & VFC & V.F. Corporation \\
Consumer Cyclical & WHR & Whirlpool Corporation \\
Consumer Cyclical & WYNN & Wynn Resorts, Limited \\
Consumer Cyclical & YUM & Yum! Brands, Inc. \\
Consumer Defensive & ADM & Archer-Daniels-Midland Company \\
Consumer Defensive & CAG & Conagra Brands, Inc. \\
Consumer Defensive & CHD & Church \& Dwight Co., Inc. \\
Consumer Defensive & CL & Colgate-Palmolive Company \\
Consumer Defensive & CLX & The Clorox Company \\
Consumer Defensive & COST & Costco Wholesale Corporation \\
Consumer Defensive & CPB & Campbell Soup Company \\
Consumer Defensive & DLTR & Dollar Tree, Inc. \\
Consumer Defensive & EL & The Estée Lauder Companies Inc. \\
Consumer Defensive & GIS & General Mills, Inc. \\
Consumer Defensive & HRL & Hormel Foods Corporation \\
Consumer Defensive & HSY & The Hershey Company \\
Consumer Defensive & K & Kellogg Company \\
Consumer Defensive & KMB & Kimberly-Clark Corporation \\
Consumer Defensive & KO & The Coca-Cola Company \\
Consumer Defensive & KR & The Kroger Co. \\
Consumer Defensive & MDLZ & Mondelez International, Inc. \\
Consumer Defensive & MKC & McCormick \& Company, Incorporated \\
Consumer Defensive & MNST & Monster Beverage Corporation \\
Consumer Defensive & MO & Altria Group, Inc. \\
Consumer Defensive & NWL & Newell Brands Inc. \\
Consumer Defensive & PEP & PepsiCo, Inc. \\
Consumer Defensive & PG & The Procter \& Gamble Company \\
Consumer Defensive & SJM & The J. M. Smucker Company \\
Consumer Defensive & STZ & Constellation Brands, Inc. \\
Consumer Defensive & SYY & Sysco Corporation \\
Consumer Defensive & TAP & Molson Coors Beverage Company \\
Consumer Defensive & TGT & Target Corporation \\
Consumer Defensive & TSN & Tyson Foods, Inc. \\
Consumer Defensive & WMT & Walmart Inc. \\
Energy & APA & APA Corporation \\
Energy & COP & ConocoPhillips \\
Energy & CVX & Chevron Corporation \\
Energy & DVN & Devon Energy Corporation \\
Energy & EOG & EOG Resources, Inc. \\
Energy & FTI & TechnipFMC plc \\
Energy & HAL & Halliburton Company \\
Energy & HES & Hess Corporation \\
Energy & HP & Helmerich \& Payne, Inc. \\
Energy & MRO & Marathon Oil Corporation \\
Energy & NOV & NOV Inc. \\
Energy & OKE & ONEOK, Inc. \\
Energy & PXD & Pioneer Natural Resources Company \\
Energy & SLB & Schlumberger Limited \\
Energy & VLO & Valero Energy Corporation \\
Energy & WMB & The Williams Companies, Inc. \\
Energy & XOM & Exxon Mobil Corporation \\
Financial Services & AFL & Aflac Incorporated \\
Financial Services & AIG & American International Group, Inc. \\
Financial Services & AIZ & Assurant, Inc. \\
Financial Services & AJG & Arthur J. Gallagher \& Co. \\
Financial Services & AMG & Affiliated Managers Group, Inc. \\
Financial Services & AMP & Ameriprise Financial, Inc. \\
Financial Services & AON & Aon plc \\
Financial Services & AXP & American Express Company \\
Financial Services & BAC & Bank of America Corporation \\
Financial Services & BEN & Franklin Resources, Inc. \\
Financial Services & BK & The Bank of New York Mellon Corporation \\
Financial Services & BLK & BlackRock, Inc. \\
Financial Services & C & Citigroup Inc. \\
Financial Services & CINF & Cincinnati Financial Corporation \\
Financial Services & CMA & Comerica Incorporated \\
Financial Services & CME & CME Group Inc. \\
Financial Services & FITB & Fifth Third Bancorp \\
Financial Services & GS & The Goldman Sachs Group, Inc. \\
Financial Services & HBAN & Huntington Bancshares Incorporated \\
Financial Services & HIG & The Hartford Financial Services Group, Inc. \\
Financial Services & ICE & Intercontinental Exchange, Inc. \\
Financial Services & IVZ & Invesco Ltd. \\
Financial Services & JPM & JPMorgan Chase \& Co. \\
Financial Services & KEY & KeyCorp \\
Financial Services & L & Loews Corporation \\
Financial Services & LNC & Lincoln National Corporation \\
Financial Services & MCO & Moody's Corporation \\
Financial Services & MET & MetLife, Inc. \\
Financial Services & MMC & Marsh \& McLennan Companies, Inc. \\
Financial Services & MS & Morgan Stanley \\
Financial Services & MTB & M\&T Bank Corporation \\
Financial Services & NDAQ & Nasdaq, Inc. \\
Financial Services & NTRS & Northern Trust Corporation \\
Financial Services & PFG & Principal Financial Group, Inc. \\
Financial Services & PGR & The Progressive Corporation \\
Financial Services & PNC & The PNC Financial Services Group, Inc. \\
Financial Services & PRU & Prudential Financial, Inc. \\
Financial Services & RF & Regions Financial Corporation \\
Financial Services & RJF & Raymond James Financial, Inc. \\
Financial Services & SCHW & The Charles Schwab Corporation \\
Financial Services & SPGI & S\&P Global Inc. \\
Financial Services & STT & State Street Corporation \\
Financial Services & TROW & T. Rowe Price Group, Inc. \\
Financial Services & TRV & The Travelers Companies, Inc. \\
Financial Services & UNM & Unum Group \\
Financial Services & USB & U.S. Bancorp \\
Financial Services & WFC & Wells Fargo \& Company \\
Financial Services & ZION & Zions Bancorporation, National Association \\
Healthcare & A & Agilent Technologies, Inc. \\
Healthcare & ABC & AmerisourceBergen Corporation \\
Healthcare & ABT & Abbott Laboratories \\
Healthcare & ALGN & Align Technology, Inc. \\
Healthcare & AMGN & Amgen Inc. \\
Healthcare & BAX & Baxter International Inc. \\
Healthcare & BDX & Becton, Dickinson and Company \\
Healthcare & BIIB & Biogen Inc. \\
Healthcare & BMY & Bristol-Myers Squibb Company \\
Healthcare & BSX & Boston Scientific Corporation \\
Healthcare & CI & The Cigna Group \\
Healthcare & CNC & Centene Corporation \\
Healthcare & COO & The Cooper Companies, Inc. \\
Healthcare & CVS & CVS Health Corporation \\
Healthcare & DGX & Quest Diagnostics Incorporated \\
Healthcare & DVA & DaVita Inc. \\
Healthcare & EW & Edwards Lifesciences Corporation \\
Healthcare & GILD & Gilead Sciences, Inc. \\
Healthcare & HOLX & Hologic, Inc. \\
Healthcare & HSIC & Henry Schein, Inc. \\
Healthcare & HUM & Humana Inc. \\
Healthcare & IDXX & IDEXX Laboratories, Inc. \\
Healthcare & ILMN & Illumina, Inc. \\
Healthcare & INCY & Incyte Corporation \\
Healthcare & ISRG & Intuitive Surgical, Inc. \\
Healthcare & JNJ & Johnson \& Johnson \\
Healthcare & LH & Laboratory Corporation of America Holdings \\
Healthcare & LLY & Eli Lilly and Company \\
Healthcare & MDT & Medtronic plc \\
Healthcare & MRK & Merck \& Co., Inc. \\
Healthcare & MTD & Mettler-Toledo International Inc. \\
Healthcare & PFE & Pfizer Inc. \\
Healthcare & PRGO & Perrigo Company plc \\
Healthcare & REGN & Regeneron Pharmaceuticals, Inc. \\
Healthcare & RMD & ResMed Inc. \\
Healthcare & SYK & Stryker Corporation \\
Healthcare & TMO & Thermo Fisher Scientific Inc. \\
Healthcare & UHS & Universal Health Services, Inc. \\
Healthcare & UNH & UnitedHealth Group Incorporated \\
Healthcare & VRTX & Vertex Pharmaceuticals Incorporated \\
Healthcare & WAT & Waters Corporation \\
Healthcare & WBA & Walgreens Boots Alliance, Inc. \\
Healthcare & XRAY & DENTSPLY SIRONA Inc. \\
Healthcare & ZBH & Zimmer Biomet Holdings, Inc. \\
Industrials & AAL & American Airlines Group Inc. \\
Industrials & ADP & Automatic Data Processing, Inc. \\
Industrials & ALK & Alaska Air Group, Inc. \\
Industrials & AME & AMETEK, Inc. \\
Industrials & AOS & A. O. Smith Corporation \\
Industrials & BA & The Boeing Company \\
Industrials & CAT & Caterpillar Inc. \\
Industrials & CHRW & C.H. Robinson Worldwide, Inc. \\
Industrials & CMI & Cummins Inc. \\
Industrials & CSX & CSX Corporation \\
Industrials & CTAS & Cintas Corporation \\
Industrials & DE & Deere \& Company \\
Industrials & DOV & Dover Corporation \\
Industrials & EFX & Equifax Inc. \\
Industrials & EMR & Emerson Electric Co. \\
Industrials & ETN & Eaton Corporation plc \\
Industrials & EXPD & Expeditors International of Washington, Inc. \\
Industrials & FAST & Fastenal Company \\
Industrials & FDX & FedEx Corporation \\
Industrials & FLS & Flowserve Corporation \\
Industrials & GD & General Dynamics Corporation \\
Industrials & GE & General Electric Company \\
Industrials & GPN & Global Payments Inc. \\
Industrials & GWW & W.W. Grainger, Inc. \\
Industrials & ITW & Illinois Tool Works Inc. \\
Industrials & JBHT & J.B. Hunt Transport Services, Inc. \\
Industrials & JCI & Johnson Controls International plc \\
Industrials & LMT & Lockheed Martin Corporation \\
Industrials & LUV & Southwest Airlines Co. \\
Industrials & MAS & Masco Corporation \\
Industrials & NOC & Northrop Grumman Corporation \\
Industrials & NSC & Norfolk Southern Corporation \\
Industrials & PAYX & Paychex, Inc. \\
Industrials & PCAR & PACCAR Inc \\
Industrials & PH & Parker-Hannifin Corporation \\
Industrials & PNR & Pentair plc \\
Industrials & PWR & Quanta Services, Inc. \\
Industrials & RHI & Robert Half Inc. \\
Industrials & ROK & Rockwell Automation, Inc. \\
Industrials & RSG & Republic Services, Inc. \\
Industrials & SWK & Stanley Black \& Decker, Inc. \\
Industrials & TXT & Textron Inc. \\
Industrials & UNP & Union Pacific Corporation \\
Industrials & UPS & United Parcel Service, Inc. \\
Industrials & URI & United Rentals, Inc. \\
Industrials & WM & Waste Management, Inc. \\
Real Estate & O & Realty Income Corporation \\
Technology & AAPL & Apple Inc. \\
Technology & ACN & Accenture plc \\
Technology & ADBE & Adobe Inc. \\
Technology & ADI & Analog Devices, Inc. \\
Technology & ADSK & Autodesk, Inc. \\
Technology & AKAM & Akamai Technologies, Inc. \\
Technology & AMAT & Applied Materials, Inc. \\
Technology & AMD & Advanced Micro Devices, Inc. \\
Technology & ANSS & ANSYS, Inc. \\
Technology & APH & Amphenol Corporation \\
Technology & CDNS & Cadence Design Systems, Inc. \\
Technology & CRM & Salesforce, Inc. \\
Technology & CSCO & Cisco Systems, Inc. \\
Technology & CTSH & Cognizant Technology Solutions Corporation \\
Technology & DXC & DXC Technology Company \\
Technology & FFIV & F5, Inc. \\
Technology & FIS & Fidelity National Information Services, Inc. \\
Technology & GLW & Corning Incorporated \\
Technology & GRMN & Garmin Ltd. \\
Technology & HPQ & HP Inc. \\
Technology & IBM & International Business Machines Corporation \\
Technology & INTC & Intel Corporation \\
Technology & INTU & Intuit Inc. \\
Technology & IT & Gartner, Inc. \\
Technology & JNPR & Juniper Networks, Inc. \\
Technology & KLAC & KLA Corporation \\
Technology & LRCX & Lam Research Corporation \\
Technology & MCHP & Microchip Technology Incorporated \\
Technology & MSFT & Microsoft Corporation \\
Technology & MSI & Motorola Solutions, Inc. \\
Technology & MU & Micron Technology, Inc. \\
Technology & NTAP & NetApp, Inc. \\
Technology & NVDA & NVIDIA Corporation \\
Technology & ORCL & Oracle Corporation \\
Technology & QCOM & QUALCOMM Incorporated \\
Technology & ROP & Roper Technologies, Inc. \\
Technology & SNPS & Synopsys, Inc. \\
Technology & STX & Seagate Technology Holdings plc \\
Technology & SWKS & Skyworks Solutions, Inc. \\
Technology & TXN & Texas Instruments Incorporated \\
Technology & VRSN & VeriSign, Inc. \\
Technology & WDC & Western Digital Corporation \\
Utilities & AEE & Ameren Corporation \\
Utilities & AEP & American Electric Power Company, Inc. \\
Utilities & AES & The AES Corporation \\
Utilities & CMS & CMS Energy Corporation \\
Utilities & CNP & CenterPoint Energy, Inc. \\
Utilities & D & Dominion Energy, Inc. \\
Utilities & DTE & DTE Energy Company \\
Utilities & DUK & Duke Energy Corporation \\
Utilities & ED & Consolidated Edison, Inc. \\
Utilities & EIX & Edison International \\
Utilities & ES & Eversource Energy \\
Utilities & ETR & Entergy Corporation \\
Utilities & EXC & Exelon Corporation \\
Utilities & FE & FirstEnergy Corp. \\
Utilities & LNT & Alliant Energy Corporation \\
Utilities & NEE & NextEra Energy, Inc. \\
Utilities & NI & NiSource Inc. \\
Utilities & NRG & NRG Energy, Inc. \\
Utilities & PEG & Public Service Enterprise Group Incorporated \\
Utilities & PNW & Pinnacle West Capital Corporation \\
Utilities & SO & The Southern Company \\
Utilities & SRE & Sempra \\
Utilities & WEC & WEC Energy Group, Inc. \\
Utilities & XEL & Xcel Energy Inc. \\
\bottomrule
%\end{tabular}
\end{longtable}}
%\end{adjustbox}

\end{document}